\documentclass[amssymb,pra,twocolumn,longbibliography,notitlepage,superscriptaddress]{revtex4-2}

\usepackage{amsmath,amssymb,amsfonts,amstext}
\usepackage{graphicx,breakurl,color,setspace}
\usepackage{mathtools}
\usepackage{dsfont}
\usepackage[T1]{fontenc}
\usepackage[latin9]{inputenc}
\usepackage[normalem]{ulem}
\usepackage[unicode=true,pdfusetitle,pdfborder={0 0 1}]{hyperref}
\usepackage{upgreek}
\usepackage{bm}
\usepackage{mathrsfs}
\usepackage{qcircuit}
\usepackage{ulem}
\usepackage{subcaption}
\let\oldcite\cite
\renewcommand{\cite}[1]{\mbox{\oldcite{#1}}}

\hypersetup{colorlinks,linkcolor=blue,citecolor=blue,filecolor=blue,urlcolor=blue}

\newcommand{\ket}[1]{| #1 \rangle} 
\newcommand{\bracket}[3]{\langle #1 | #2 | #3 \rangle} 

\newcommand{\etal}{\textit{et al.}}
\newcommand{\vect}[1]{\boldsymbol{#1}}

\newcommand{\vgamma}{\vec{\bm\gamma}}
\newcommand{\vbeta}{\vec{\bm\beta}}

\begin{document}

\title{A loop Quantum Approximate Optimization Algorithm with Hamiltonian updating}

\author{Fang-Gang Duan}
\affiliation{Guangdong Provincial Key Laboratory of Quantum Engineering and Quantum Materials, School of Physics and Telecommunication Engineering, South China Normal University, Guangzhou 510006, China}
\author{Dan-Bo Zhang}
\email{dbzhang@m.scnu.edu.cn}
\affiliation{Guangdong Provincial Key Laboratory of Quantum Engineering and Quantum Materials, School of Physics and Telecommunication Engineering, South China Normal University, Guangzhou 510006, China}
\affiliation{Guangdong-Hong Kong Joint Laboratory of Quantum Matter, Frontier Research Institute for Physics, South China Normal University, Guangzhou 510006,
	China}

\date{\today}

\begin{abstract}
Designing noisy-resilience quantum algorithms is indispensable for practical applications on Noisy Intermediate-Scale Quantum~(NISQ) devices. Here we propose a quantum approximate optimization algorithm~(QAOA) with a very shallow circuit, called loop-QAOA, to avoid issues of noises at intermediate depths, while still can be able to exploit the power of quantum computing. The key point is to use outputs of shallow-circuit QAOA as a bias to update the problem Hamiltonian that encodes the solution as the ground state. By iterating a loop between updating the problem Hamiltonian  and optimizing the parameterized quantum circuit, the loop-QAOA can gradually transform the problem Hamiltonian to one easy for solving. We demonstrate the loop-QAOA on Max-Cut problems both with and without noises. Compared with the conventional QAOA whose performance will decrease due to noises, the performance of the loop-QAOA can still get better with an increase in the number of loops. The insight of exploiting outputs from shallow circuits as bias may be applied for other quantum algorithms. 
 
\end{abstract}

\maketitle

\section{Introduction}
Recent advances on quantum computing technology advocate a growing interest in finding quantum algorithms that can be implemented effectively on NISQ devices~\cite{Preskill2018quantumcomputingin}. One such strong candidate is variational quantum algorithm~(VQA)~\cite{cerezo_variational_2021,McClean2016,Kandala2017,McArdle2019,Moll_2018}, which leverages up the expressive power of parameterized quantum circuits and 
a hybrid quantum-classical optimization procedure for finding solutions. Variational quantum algorithms have wide applications for solving quantum many-body systems~\cite{liu_19,kokail_19}, quantum chemistry~\cite{RevModPhys_yuanxiao,Peruzzo2014,OMalley2016,yuan_PRA_2021,google_quantum_chemistry}, optimization problems~\cite{farhi2014quantum,zhou_PRX_2020,harrigan_quantum_2021,McClean_PRXQuantum_2021}, and so on.  Among them, the quantum approximate optimization algorithm~(QAOA) stands out as a VQA for tackling hard combinatorial optimization problems~\cite{farhi2015quantum,Wecker2016Training,YangVQA,Ho2018Efficient,QAOAGrover,Anschuetz2018Variational,Moussa_2020,Akshay_PRL_2020,Bravyi_PRL_2020}.

The QAOA applies an alternating evolution of mixer and problem Hamiltonians with optimized evolution periods that its outputs contain the target solution with a large weighting. By exploiting quantum effects such as interference, QAOA is believed to show quantum advantages on combinatorial optimization problems at intermediate depths~\cite{farhi2019quantum,zhou_PRX_2020}. One center challenge for implementing QAOA, however, is that noises on near-term quantum processors will be detrimental to the performance. For instance, the recent research of the Google team shows that QAOA performs worse at depths larger than $p=3$~\cite{harrigan_quantum_2021}. Moreover, the non-convex and high-dimensional parameter landscape caused by large $p$ may bring huge difficulty for optimizing~\cite{Bittel_PRL_2021,Shaydulin_2019}. Without efficient approaches for optimization, the hybrid algorithms will lose it's potential advatages~\cite{McCleanBarrenLandscape}. 

Those challenges addressed at intermediate depths of QAOA have being tackled recently, e.g., the issue of noise can be reduced by error mitigation techniques~\cite{temme_error_2017,Endo_2021}, and some constructive optimization procedures have been proposed~\cite{zhou_PRX_2020,Akshay_PhysRevA_2021}. An alternative approach, however, is to ask whether we can exploit very shallow QAOA to achieve good performance under noise. As the circuit depth is very shallow, there is no difficulty of variational parameter optimization, and effects of noises can also be small. Remarkably, even the lowest depth~($p=1$) of QAOA has been proved that classical computers can't simulate efficiently~\cite{farhi2019quantum}. This suggests a direction for designing noise-resilience QAOA using shallow circuits while the quantum computing power may  still be exploited.

Our approach is to explore both the parameterized quantum circuit and the problem Hamiltonian in QAOA for finding the target solution. In the conventional QAOA, the Hamiltonian which encodes the problem is usually fixed. However, it is possible that many different Hamiltonians can have the same solutions or approximated solutions but may be solved with QAOA at different degrees of difficulty. Motivated by this, we may first use a shallow QAOA to get some rough estimations of the solution, and then update the problem Hamiltonian with biased terms. By iterating the procedure of updating the problem Hamiltonian and finding the optimized parameters of the quantum circuit, there is a feedback loop between those two subroutines and the solution may get better with an increase of loops. In this regard, we may call it as \emph{loop-QAOA}. 

In this work, we propose a kind of noise-resilience quantum approximate optimization algorithm, the loop-QAOA, for solving combinatorial optimization problems with very shallow circuits, with a feedback loop between Hamiltonian updating and parameter optimization. Different strategies for constructing bias field terms are utilized in the Hamiltonian updating. We test the loop-QAOA and compare it to conventional QAOA on typical Max-Cut graph instances up to 12-vertex instances, simulated both with and without noise. The results suggest that loop-QAOA can be a practical quantum algorithm on noisy quantum devices, where the conventional QAOA may fail to improve its performance by increasing the circuit depth.
Our work gives a new sight for designing hybrid quantum-classical algorithms using feedback from shallow quantum circuits.   
 
The paper is organized as follows:  we firstly review QAOA and the Max-Cut problem in Sec.~\ref{Sec:background}. Then we will propose loop-QAOA and some strategies for constructing bias field terms in Sec.~\ref{Sec:loop-QAOA}. The simulated results are shown in Sec.~\ref{Sec:Performance}. Finally, conclusions are given in Sec.~\ref{conclusions}. 


\section{Quantum Approximate Optimization Algorithm revisited} \label{Sec:background}
The Quantum Approximate Optimization Algorithm (QAOA) by Farhi et al. is a hybrid quantum-classical algorithm{, designed to produce approximated solutions for combinatorial optimization problems. The solutions are encoded as the ground state of $H_P$, the problem Hamiltonian or  the phase Hamiltonian. 
 The $H_P$ is diagonal in the computational basis. As various hard combinatorial optimization problems can be encoded in such diagonal Hamiltonians~\cite{Lucas_2014}, the QAOA is often considered as a potential method of solving combinatorial optimization problems on quantum computers. These problems are usually defined on $N$-bit binary strings $\vect {z}$ = $z_\text {1}z_\text {2}\cdots z_N$. The goal is to find a string which maximizes the  classical objective functions $C(\vect{z}):\{+1,-1\}^{\otimes N} \mapsto\mathds{R}$. 

As the first step of the QAOA, the classical objective function is encoded in a quantum Hamiltonian $H_P$ defined on $N$-qubits
by replacing each variable $z_i$ with a single-qubit $\sigma_i^z$. The ground state of $H_P$ is parametrized with a quantum circuit which is constructed as follows. For a $p$-layer QAOA, the state is initialized as $\ket{+}^{\otimes N}$, where $\ket{+}=(\ket 0+\ket 1)/\sqrt {2}$. Let $( \vbeta,\vgamma)=(\beta_1,\cdots \beta_p,\gamma_1\cdots,\gamma_ p)$ be real postive numbers for the layer $p$.  The phase Hamiltonian $H_P$ and a mixer Hamiltonian, often chosen as $H_M=\sum_i\sigma_i^x$, are then applied alternately to generate a variational wavefunction
\begin{equation}\label{eq1}
\ket{\Psi_{p}(\vbeta,\vgamma)} = \prod_{j=1}^p e^{-i \beta_{j} H_{M}} e^{-i \gamma_{j} H_{P}}  \ket{+}^{\otimes N}.
\end{equation}
 The expectation value $H_P$ with regard to the variational state
is 
\begin{equation}\label{eq2}
F_p(\vgamma, \vbeta) = \bracket{\Psi_{p}(\vbeta,\vgamma)}{H_{P}}{\Psi_{p}(\vbeta,\vgamma)},
\end{equation}
which can be obtained by measurements in the computational basis.
The goal is to make 
$F_{p}(\vgamma, \vbeta)$ minimal. The parameters $\beta_1,\cdots \beta_p,\gamma_1\cdots,\gamma_ p$ are adjusted in repeated learning loops in a classical computer to minimize the objective function Eq.~\eqref{eq2} evaluated with a quantum computer.

We use the Max-Cut to illustrate the QAOA more concisely. Consider a graph $G$ with $N$ vertices over a vertex set $V$ and edge set $E$. The Max-Cut problem is to divide the vertexes $V$ into two disjoint subsets such that the number of cuts~(edges connecting two disjoint subsets) is maximized.
Let $\omega_{ij}$ be the weight of the associated edge $\left<i,j \right>\in E$. The objective function can be defined as a sum of edge terms in the graph as
\begin{equation}
C(\vect{z})=\sum_{\left<i,j \right>\in E} C_{ij}=\frac {1}{2}\sum_{\left<i,j \right>\in E}  \omega_{ij}(1-z_iz_j)
\end{equation}
where $z_i=\pm 1$. By maximizing $C(\vect{z}$, the cut set $S(z)\in E$ can be obtained for the Max-Cut problem. 

The Max-Cut Problem is in the APX-complete complexity class, meaning that there is no polynomial-time approximation scheme to find the optimal solution, unless P = NP~\cite{Hastad2001,Berman1999,PAPADIMITRIOU1991425}. But we may find a polynomial-time approximation algorithm with a solution of a string $\vect {z'}$ that achieves a desired approximation ratio
\begin{equation}
\frac {C(\vect {z'})}{\max\{C(\vect {z})\}}\ge r^*.
\end{equation}


The objective function of the Max-Cut problem can be encoded as a problem Hamiltonian, 
\begin{equation}
H_P=-\frac {1}{2}\sum_{\left<i,j \right>\in E}  \omega_{ij}(1-\sigma_i^z\sigma_j^z).
\end{equation}
Note an addition minus sign in $H_P$ turns the task of maximizing the objective function $C(\vect{z})$ to finding the lowest energy of $H_P$. 
If spins $\sigma_i^z$ and $\sigma_j^z$ are anti-aligned, an edge $\left<i,j \right>$ contributes with an energy $-\omega_{ij}$; otherwise, the energy contribution is zero. 

It has been proven NP-hard to design an algorithm that guarantees an approximation ratio of $r^*\ge 16/17\approx 0.941$ for Max-Cut on all graphs~\cite{Hastad2001}. The best-known classical approximation algorithm, of Goemans and Williamson (GW)~\cite{GW}, guarantees an approximation ratio of $r^*\approx 0.878$ using semidefinite programming and randomized rounding. For more special graphs the GW can do better, $r^*$ can reach 0.932 for u3R graphs~\cite{Halperin2004}. Farhi \etal~\cite{farhi2014quantum} showed that QAOA at level $p=1$ the worst-case approximation ratio is 0.6924 for u3R graphs.  It is known the performance of QAOA gets better with an increase of the depth $p$~\cite{farhi2014quantum}. The goal of quantum approximate optimization is to raise $r^*$ that is beyond the capacity of classical algorithms.

The QAOA often requires quantum circuits of intermediate depths to achieve the desired accuracy for combinatorial optimization problems. However, the benefit of increasing $p$ in QAOA will be hampered due to noises in quantum processors. The performance in the presence of noises may get worse for $p$ larger than a certain small value~\cite{harrigan_quantum_2021}.  Even without noise, optimization of variational parameters for large $p$ can be a great challenge. The two issues can be avoided if a shallow circuit is used in QAOA, which would be explored in the next section. 

\vspace{-4pt}
\section{Loop-QAOA with problem Hamiltonian updating}
\label{Sec:loop-QAOA}
\vspace{-4pt}

In QAOA, it is both the problem Hamiltonian and the parameterized quantum circuit to determine the solution, as seen in Eq.~\eqref{eq2}. Usually, the problem Hamiltonian is given beforehand, and what is left is to increase the depth $p$ or design 
novel variational ansatz. Nevertheless, there is plenty of room for handling the problem Hamiltonian. Since the same solution may correspond to different problem Hamiltonians. It is possible that the problem Hamiltonian can be transformed to a new one with the same ground state but easier to solve. Instead of increasing the depth $p$, one can fix QAOA with a very small $p$, while the problem Hamiltonian is improved. Guided by this insight, we propose the loop-QAOA, where the problem Hamiltonian is updated in an iterative fashion with the feedback from shallow-circuit QAOA.  


Let us first elaborate on the case of one loop to illustrate the mechanism behind loop-QAOA. Throughout the paper, we use the least shallow circuit in loop-QAOA, namely $p=1$, while the case $p>1$ can be investigated similarly. In practice, the $p=1$ QAOA can show much better performance than randomly guessing. Moreover, it gives a probability distribution over all possible bitstrings. The output of $p=1$ QAOA thus can reveal the underlying structure of the problem to some degree, which can be exploited as a bias for the combinatorial optimization problems. For instance, the bitstrings with high probabilities are informative for constructing bias terms for updating the problem Hamiltonian.
	

The procedure of training $p=1$ QAOA and updating the problem Hamiltonian can be iterated multiple times. Schematic of the loop-QAOA can be found in Fig.~\ref{sketch}a. The aim is to gradually transform the Hamiltonian that the chance to get the target solution will increase with more loops. To make this possible, the bias terms and the scheme of Hamiltonian updating should be designed properly. We note that using bias terms to accelerate the quantum optimization has been considered in the context of quantum annealing~\cite{Grass_PRL_2019}. 

\begin{figure}[t!]
	\centering
	\begin{subfigure}[b]{0.5\textwidth}
		\centering
		\includegraphics[width=\textwidth]{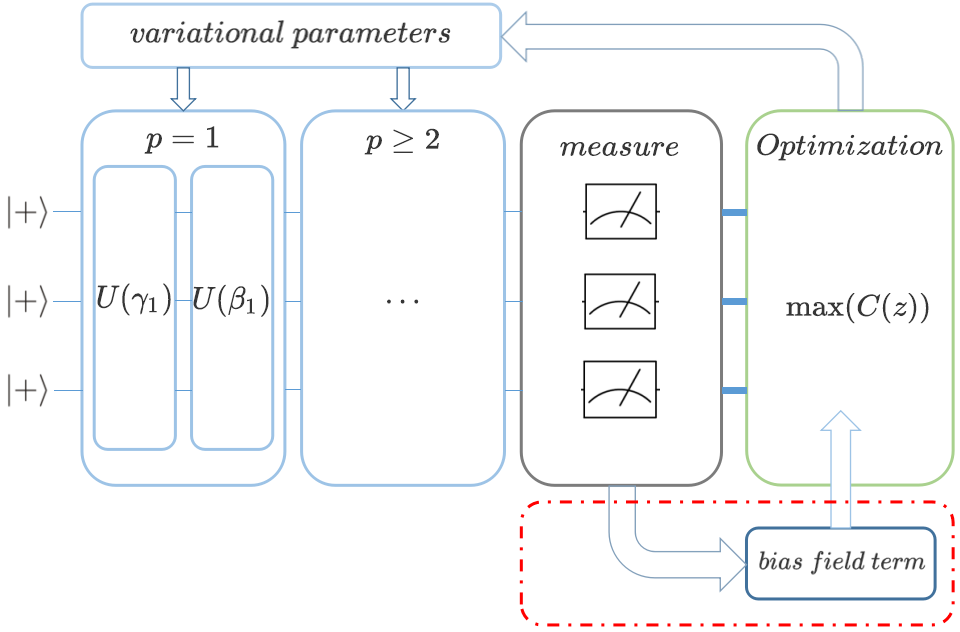}
		\caption{}
		\label{fig:a}
	\end{subfigure}
	\begin{subfigure}[b]{0.5\textwidth}
		\centering
		\includegraphics[width=\textwidth]{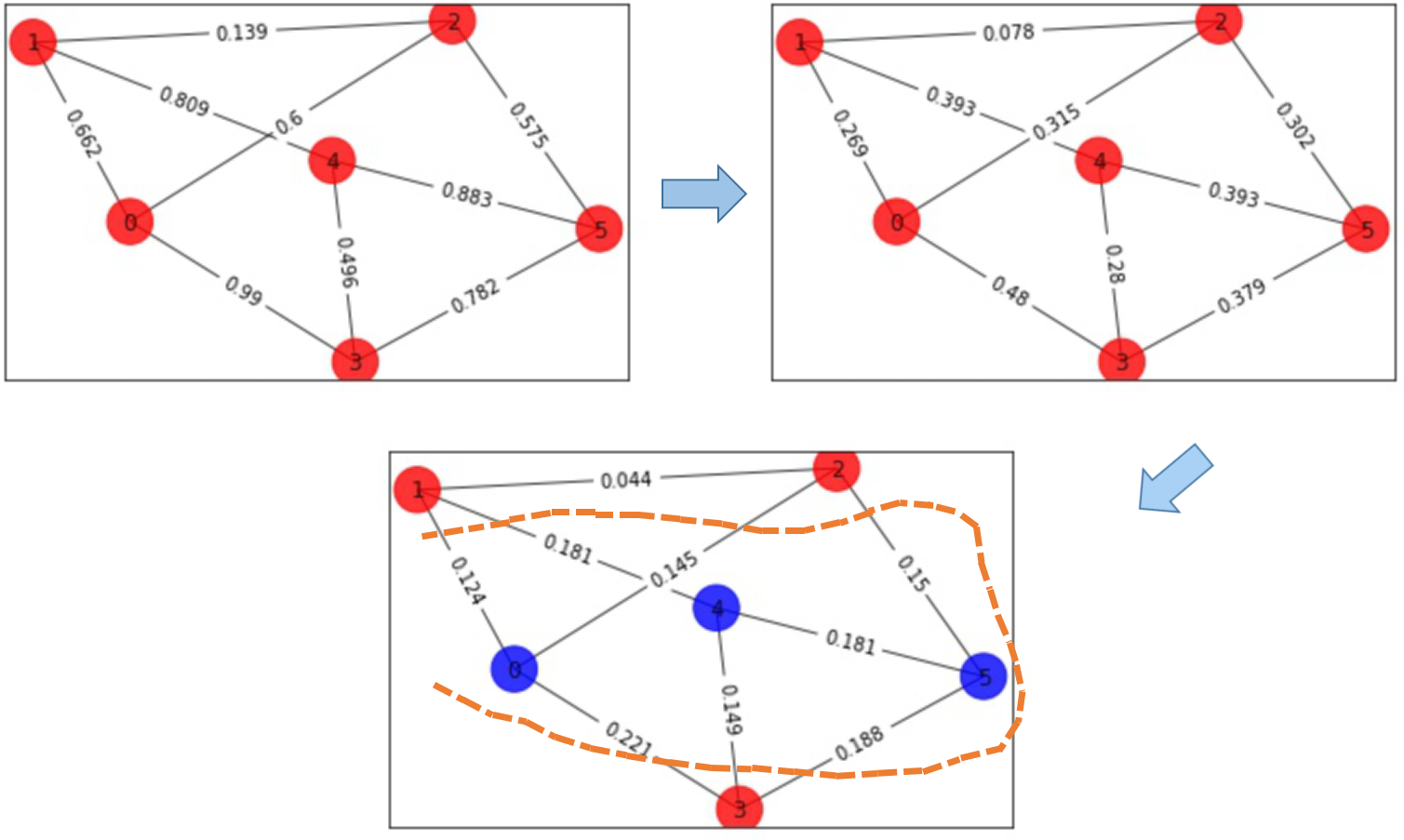}
		\caption{}
		\label{fig:b}
	\end{subfigure}
	\caption{\textbf{Schematic of a $p$-level QAOA and loop-QAOA strategy.} (a) The circuit diagram in the solid frame represents QAOA. A quantum circuit takes input $\ket{+}^{\otimes N}$ and alternately applies $U(\beta_i)=e^{-i\beta_i H_P}$ and $U(\gamma_i)=e^{-i\gamma_i H_M}$, and measure the final state to obtain expectation value with respect to the objective function $C(z)$. This result is optimized by a classical optimizer to get the maximum value. The dashed box represents the loop-QAOA strategy. We add bias field terms to objective Hamiltonian through the measurement results of each loop. (b) Schematic of loop-QAOA strategy applied to a 6-vertex w3R graph, where weights are updated at each loop.}
	\label{sketch}
\end{figure}

We illustrate the procedure of Hamiltonian updating with bias terms with the Max-Cut problem. Suppose we have a connected weighted graph $G=(V,E)$ with $N$ nodes, and the weight set of its edges is $W = \{\omega_{ij}:\left<i,j \right>\in E\}$. The output of $p=1$ QAOA is a probability distribution $Q = \{(\vect z_i, p_i)\}$, where $\vect z_i$ is the bitstring and $p_i$ is the corresponding probability. We will choose the results with $p_i$ greater than a certain threshold $p^*$, which are informative for constructing bias terms.

The construction of bias terms is based on how the distribution of bitstrings output from QAOA contributes to the objective function.
Consider a string $\vect z_1$ for instance. If two of the spins are opposite, $\vect z_{1i}=0, \vect z_{1j}=1 $ or $ \vect z_{1i}=1, \vect z_{1j}=0$ (based on the $\mathds Z_2$ symmetry of the Max-Cut problem), then the $\left<i,j \right>$ edge will make a contribution of $\omega_{ij}$ to the objective function. If the spins are the same, namely $\vect z_{1i}=\vect z_{1j}$, the contribution is $0$. Considering the distribution from measurement results, $\left<i,j \right>$ in $z_1$ makes a contribution of $p_{z_1}\omega_{ij}$. Then the contribution of the edge $\left<i,j \right>$ to the objective function in total is
\begin{equation}
C_{ij} = \sum_{k=1}^N p_{z_k}\omega_{ij}
\end{equation}
After going through all the edges in $G$, we get a set $\{C_{ij}\}$. We can reward the edge with a larger contribution and punish the edge with a smaller contribution.

The reward or punishment is realized by adding bias terms to the problem Hamiltonian. For this, we introduce a mapping $F:W\mapsto W'$, which tunes the weights of the graph. As a schematic, Fig.~\ref{sketch}b shows how weights of a 6-nodes w3R graph are updated in the loop-QAOA, which finally becomes easy to solve. There can be different mapping schemes. Apparently, different mapping methods or different mapping parameters will lead to different results. 
Here, we list some bias field methods: (1). Bias field and antibias field: $W'=\{\omega_{ij}\pm kC_{ij}\}$($k$ can be a constant or a function, for controling the bias field terms.); (2). Fixed bias field: $W'=\{\omega_{ij}+ \text {fixed bias terms}\}$. We adopt a modified version of the first scheme with bias fields that updates edge weight in each loop via the following mapping,  
\begin{equation}
\omega_{ij}'=(1-\tau\sum^N_{z_{ki}=z_{kj}}{p_{\vect{z}_k}})\omega_{ij},
\end{equation}
where $\tau$ is a parameter to control the strength of bias.

Using this scheme, the solution space will be reduced by punishing a bitstring $\vect{z}_k$ if there are many pairs of spins that are the same $z_{ki}=z_{kj}$ on the edges. Note that the weights may approach zeroes for a large number of loops. To avoid this situation, we can control the bias strength $\tau$ to make the strategy get the desired results with a small number of loops.  A heuristic formula of bias strength is proposed by interpolation:
\begin{equation}\label{eq:interpolation}
\tau = \frac {1}{2^{n-2}}[1-\frac {1}{3}(\frac {1}{g\Delta_{max}})^{\frac {1}{l+f(n)}}-\frac {2}{3}(\frac {1}{g\Delta_{min}})^{\frac {1}{l+f(n)}}]
\end{equation}
where $n$ is nodes number, $l$ is the current number of loops, $f(n)$ is a function monotonous with the number of nodes, $\Delta_{max}$ is the maximum weight difference, and $\Delta_{min}>0$ is the minimum weight difference. The factor $g$ ensures that the results are within the given accuracy range, and we chose $g=10^3$ to keep three significant digits. The power correspond to current loops of algorithm, where $f(n)$ is usually taken as a fixed constant. $\Delta_{max}$ and $\Delta_{min}$ ensure that all edges will not become too small within a certain number of loops. The coefficinet $1/3$ and $2/3$ are obtained by interpolating, and can be adjusted to other numbers.
We mention that the above method is not only limited to the Max-Cut problem but also effective for general combinatorial optimization problems.

\begin{figure}
    \centering
    \begin{subfigure}[b]{0.5\textwidth}
           \centering
           \includegraphics[width=\textwidth]{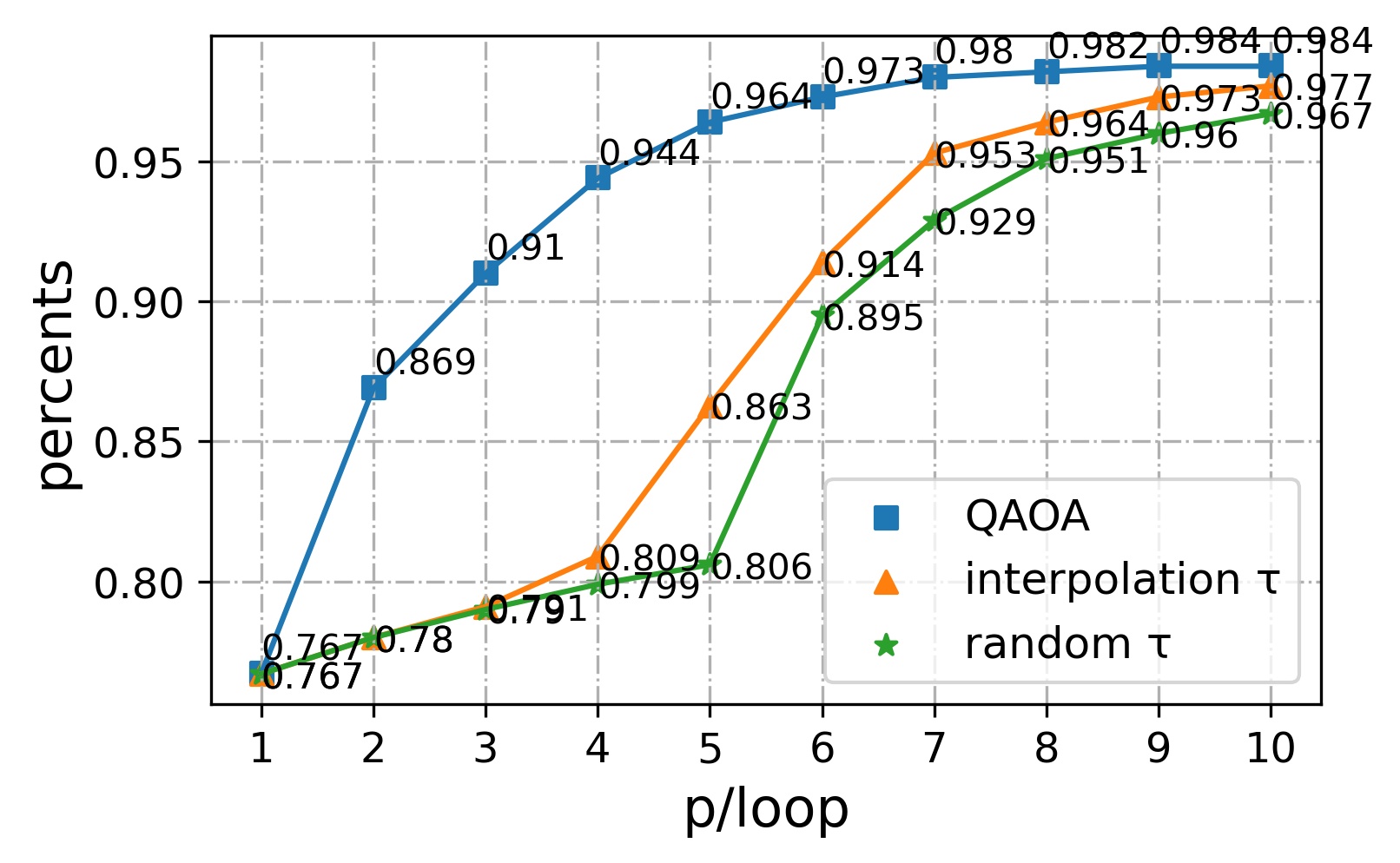}
            \caption{}
            \label{fig:a}
    \end{subfigure}
    \begin{subfigure}[b]{0.5\textwidth}
            \centering
            \includegraphics[width=\textwidth]{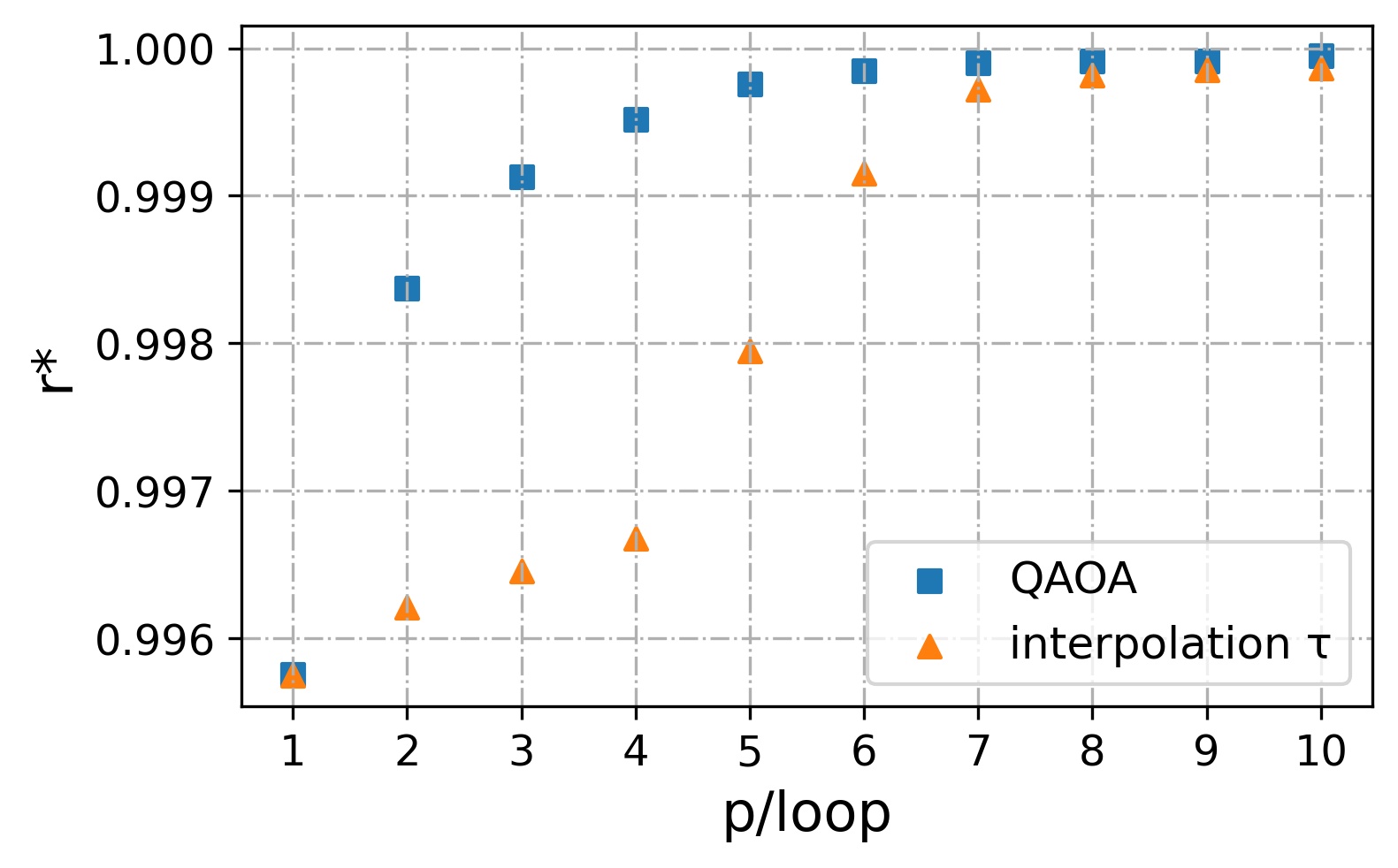}
            \caption{}
            \label{fig:b}
    \end{subfigure}
\caption{\textbf{Performances of the conventional QAOA and the loop-QAOA with two different choices of bias strengths.} (a) the average success rates $p_s$ on $10^4$ randomly generated 12-vertex w3R graphs. $\tau$ indicates the bias strength, and they are chosen by the interpolation in Eq.~\eqref{eq:interpolation}, or randomly choosing form $2.0\times 10^{-4}$ to $2.8\times 10^{-4}$. For all cases loop-QAOA gradually approaches the conventional QAOA with the increase of the number of loops. (b) The average approximation ratio $r^*$ of the QAOA and the loop-QAOA with the interpolation method. }
\label{result1}
\end{figure}


\section{Simulation Results}
\label{Sec:Performance}
In this section, we first give the simulation methods and then present the simulation results. Comparisons between the performances of loop-QAOA and the conventional QAOA are given, both in the presence or absence of noises.

\subsection{Simulation methods}
Without losing the difficulty of the combinatorial optimization problem, we are concerned about connected graphs with weights~(such as w3R) $w_{ij}$ chosen uniformly at random from the interval $[0,1]$. We consider randomly generated instances w3R graphs with vertex number $8\le N\le 14$. For each vertex number,  $10^4$ w3R graph instances are generated. The loop-QAOA strategy at each loop and the conventional QAOA at each layer $p$ are applied for those generated graphs. As for the optimization method, we apply both the gradient-based optimization routine (BFGS) in our numerical simulation. It is noted that non-gradient-based routines such as Nelder-Mead~\cite{verdon2019quantum,2017arXiv170101450G} and Bayesian methods~\cite{otterbach2017unsupervised} also works for the loop-QAOA.

We also add noise in our numerical simulation to address that the loop-QAOA can be noise-resilience. We choose three quantum noise models to test the algorithm, namely quantum bit flip channel, quantum phase flip channel, and depolarizing channel~\cite{Wilde1,nielsen_chuang_2010}. 
The quantum bit flip channel evolves a density matrix via:
\begin{equation}\label{eq:noise1}
\rho \to (1-q)\rho + q X\rho X,
\end{equation}
where $q$ is the probability of a bit flip~($q$ instead of $p$ is to distinguish it from the previous $p$-layer). The quantum phase channel evolves a density matrix via:
\begin{equation}\label{eq:noise2}
\rho \to M_0\rho M_0^{\dagger}+M_1\rho M_1^{\dagger}
\end{equation}
with :
\begin{equation}\label{eq:noise}
M_0=\sqrt{q}\begin{bmatrix} 1 & 0 \\ 0 & 1 \end{bmatrix}\quad,
M_1=\sqrt{1-q}\begin{bmatrix} 1 & 0 \\ 0 & -1 \end{bmatrix}\quad,
\end{equation}
where $q$ is the probability of a phase flip. The depolarizing channel applies one of $4^n$ disjoint possibilities: the identity channel or one of the $4^n - 1$ pauli gates. The  probability of a non-identity Pauli gates are $q/(4^n - 1)$, and the identity is done with probability $1 - q$, where $q$ is the probability that one of the Pauli gates is applied, $n$ is the number of qubits, $P_i$ are the $4^n-1$ Pauli gates(excluding the identity). This channel evolves a density matrix via
\begin{equation}\label{eq:noise3}
\rho \to (1-q)\rho + \frac {q}{4^n-1}\sum_i P_i \rho P_i
\end{equation}

We apply these models to the whole quantum circuit. We show results for the case of $q=0.1$ in bit flip channel and phase flip channel, $q=0.05$.in depolarizing channel in Fig~\ref{resultnoise}.

\begin{figure}
    \centering
    \begin{subfigure}[b]{0.5\textwidth}
           \centering
           \includegraphics[width=\textwidth]{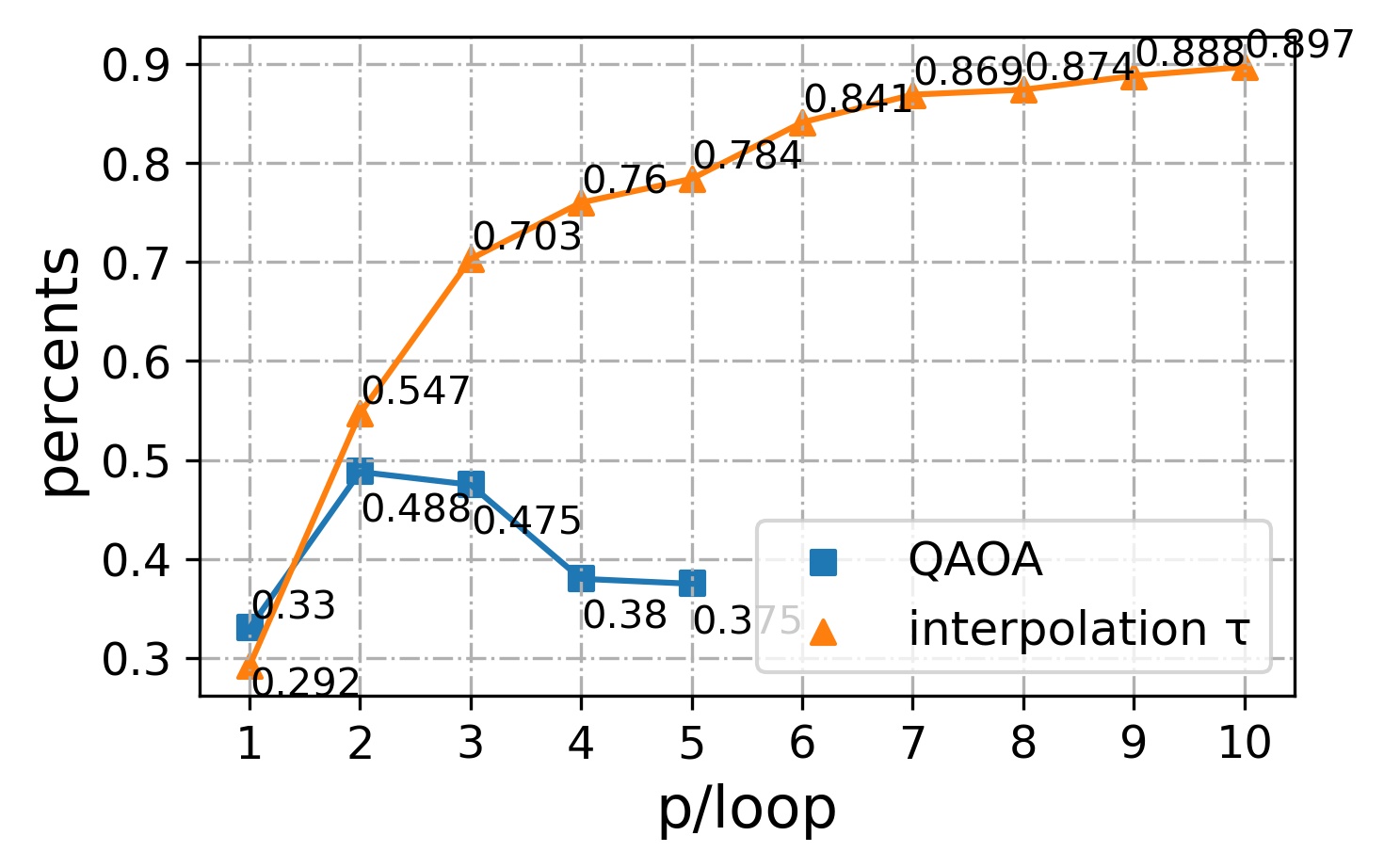}
            \caption{Bit flip channel}
            \label{fig:a}
    \end{subfigure}
    \begin{subfigure}[b]{0.5\textwidth}
            \centering
            \includegraphics[width=\textwidth]{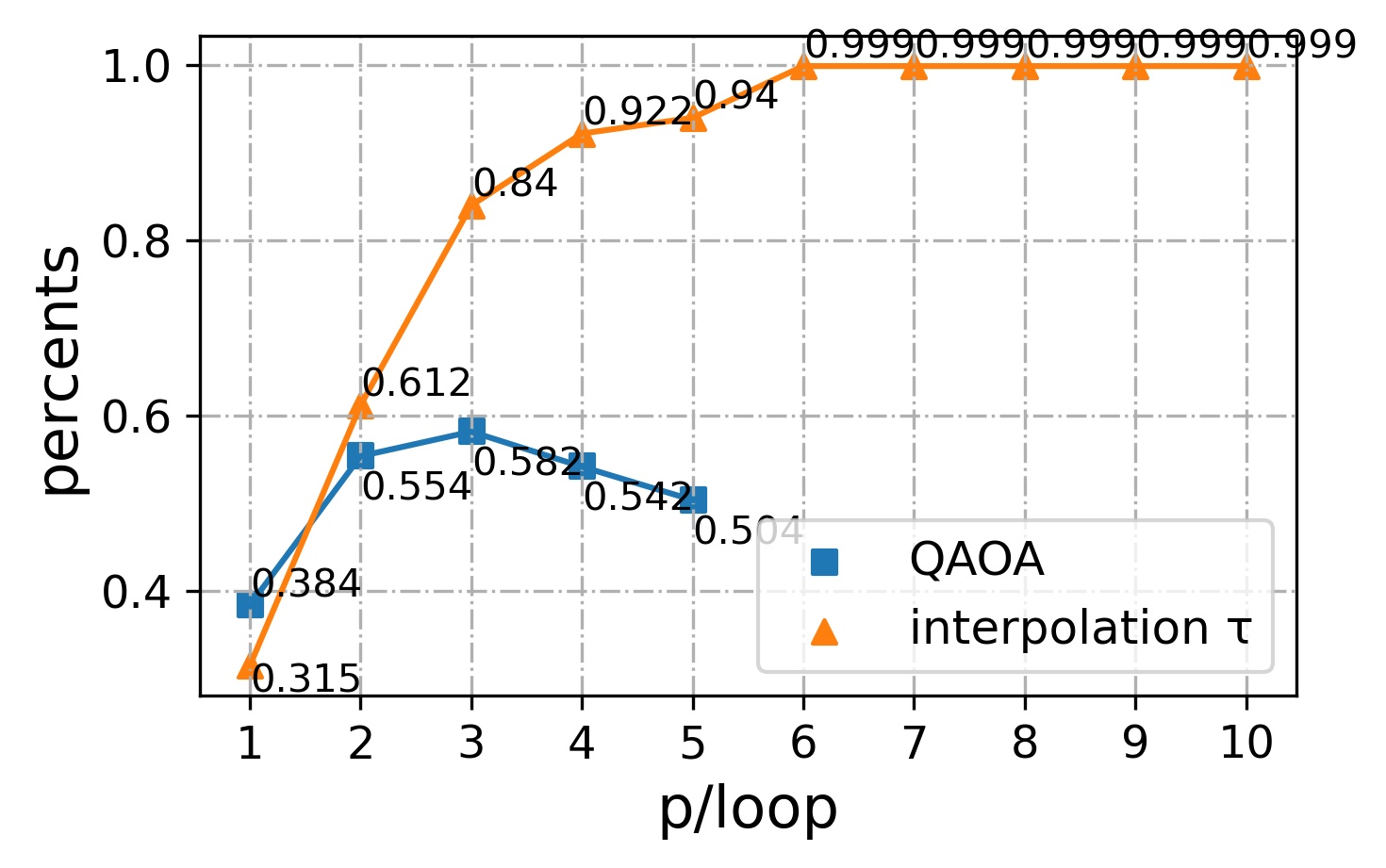}
            \caption{Phase flip channel}
            \label{fig:b}
    \end{subfigure}
    \begin{subfigure}[b]{0.5\textwidth}
            \centering
            \includegraphics[width=\textwidth]{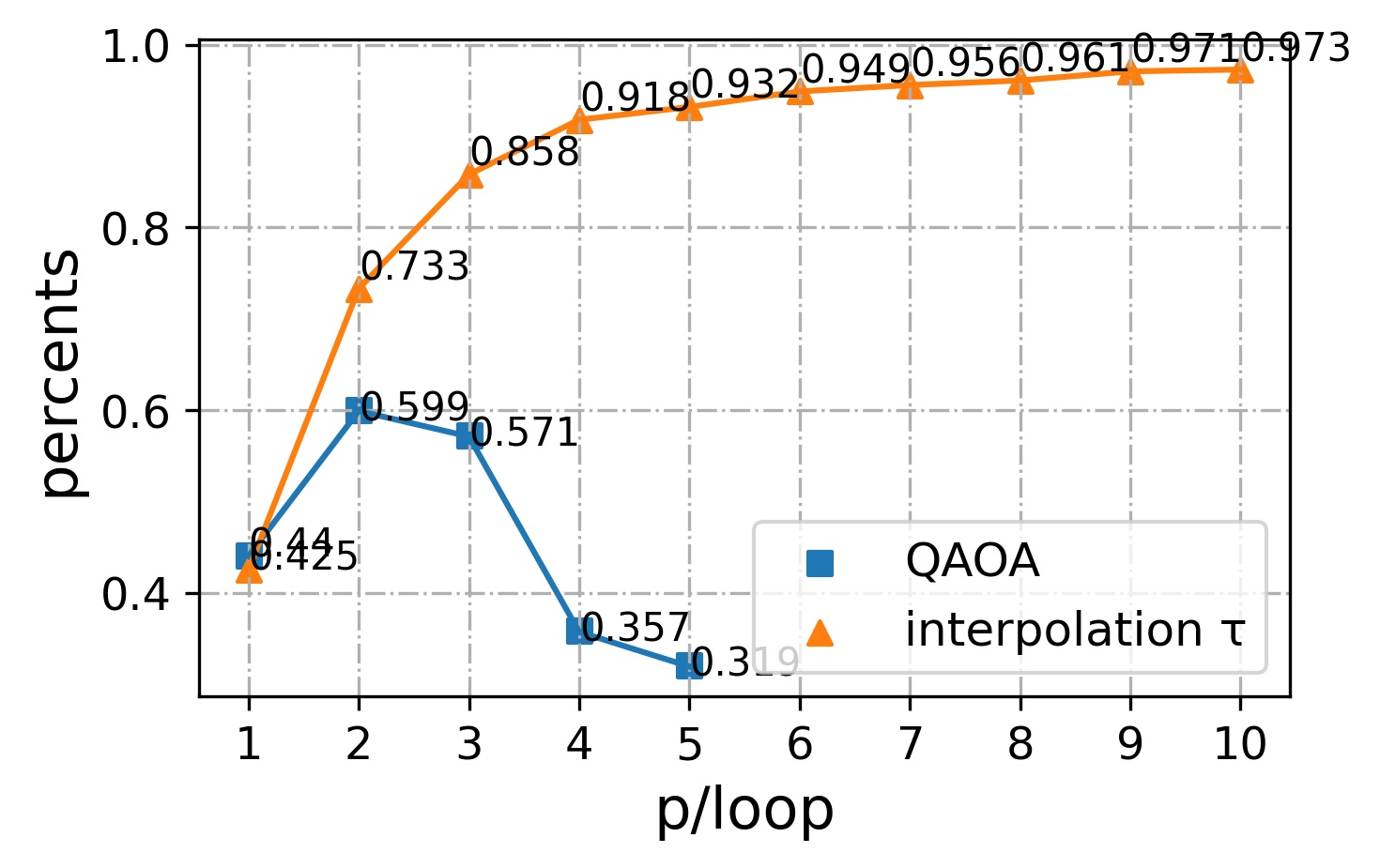}
            \caption{Depolarizing channel}
            \label{fig:c}
    \end{subfigure}
\caption{\textbf{Performances of the conventional QAOA and the loop-QAOA at three different noise model environment.} (a),(b) and (c) shows the average success rates $p_s$ on $10^4$ randomly generated 12-vertex w3R graphs with bit flip channel, phase flip channel and depolarizing channel, respectively.} 
\label{resultnoise}
\end{figure}
\subsection{Results}
We first compare the performances of the conventional QAOA and the loop-QAOA in the absence of noises. For the loop-QAOA, the bias strengths at different loops are chosen according to the interpolation in Eq.~\eqref{eq:interpolation} or chosen randomly from a given interval. 
The results on a set of 12-vertex w3R graphs with $10^4$ instances are shown in Fig.~\ref{result1}. 
Since QAOA can be regarded as the case where the bias constant $\tau=0$, the performance of the two is almost the same here at $p=1$ and loop$\ =1$. At low $p$ or low loops, the loop-QAOA strategy is slightly inferior to the original QAOA method. However, after a certain number of loops, the performance of loop-QAOA can gradually approach the original QAOA. When loop $\ge8$, the performance of the two is very close. For the conventional QAOA, this method yields a success rate of $p_s=0.983$ at $p=10$. The interpolation loop-QAOA reaches a success rate of $p_s=0.979$, which is very close to the former. On the other hand, the loop-QAOA with random choices of bias strengths performs worse than the interpolation one. The above results indicate that the loop-QAOA can be a potential powerful quantum approximate optimizer using only a very shallow quantum circuit as long as appropriate bias field parameters are selected.

Remarkably, the loop-QAOA can have an advantage over the conventional QAOA when noises should be considered. Under three noise models as described in Eq.~\eqref{eq:noise1}~\eqref{eq:noise2}~\eqref{eq:noise3}, we simulate the QAOA and the loop-QAOA on randomly generated $10^3$ instances of w3R graphs with vertex number $N=4$, and the results are shown in Fig.~\ref{resultnoise}. We limit $p$ to be less than or equal to 5 for consideration of simulation time and cost. Due to the existence of noise, the performance of QAOA improves slower with the increase of $p$ and even can decrease for larger $p$. However, the performance of the loop-QAOA still raises with the loops and finally reaches a very high success rate, which significantly outperforms the best performance of the conventional QAOA. 

\section{Conclusion}\label{conclusions}
To conclude, we have proposed the loop-QAOA as a noise-resilience quantum  algorithm for solving combinatorial optimization problems. The loop-QAOA exploits the outputs from a very shallow circuit as a bias to update the problem Hamiltonian and forms a feedback loop between Hamiltonian updating and parameter optimization. We have tested the loop-QAOA  and compared it to the conventional QAOA on the Max-Cut problem with and without noises. Remarkably, the performance of the loop-QAOA under noises can still be made better by increasing the loop number, while the performance of the conventional QAOA will cease to increase at a small depth. 
Our work points out a direction for designing quantum algorithms with practical applications on NISQ devices with very shallow circuits. 	

\begin{acknowledgments}
This work was supported by the National Natural Science Foundation of China (Grant No.12005065), by the Guangdong Basic and Applied Basic Research Fund (Grant No.2021A1515010317), and the Key Project of Science and Technology of Guangzhou(Grant No.2019050001).
\end{acknowledgments}

\bibliography{QAOArefs}
\end{document}